\documentclass[%
 reprint,
 amsmath,amssymb,
 aps,prl
]{revtex4-1}

\usepackage{graphicx}% Include figure files
\usepackage{dcolumn}% Align table columns on decimal point
\usepackage{bm}% bold math
\usepackage[hidelinks]{hyperref} % add hypertext capabilities
\begin{document}

\preprint{APS/123-QED}

\title{Strongly Anisotropic Charge Dynamics in La$_3$Ni$_2$O$_7$ with\\ Coherent-to-Incoherent Crossover of Interlayer Charge Dynamics}%
\author{Bo Su,$^{1,\dagger}$ Chaoxin Huang,$^{2,\dagger}$ Jianzhou Zhao,$^{3,\dagger}$ Mengwu Huo,$^{2}$ Jianlin Luo,$^{1}$ Meng Wang,$^{2}$ and  Zhi-Guo Chen,$^{1,4,}$} 
   \email{zgchen@iphy.ac.cn}
\affiliation{$^1$Beijing National Laboratory for Condensed Matter Physics, Institute of Physics, Chinese Academy of Sciences, Beijing 100190, China}
\affiliation{$^2$Guangdong Provincial Key Laboratory of Magnetoelectric Physics and Devices, School of Physics, Sun Yat-Sen University, Guangzhou 510275, China}
\affiliation{$^3$Co-Innovation Center for New Energetic Materials, Southwest University of Science and Technology, Mianyang Sichuan 621010, China}
\affiliation{$^4$Songshan Lake Materials Laboratory, Dongguan Guangdong 523808, China}

\begin{abstract}
We report an optical spectroscopy study of the charge-dynamics anisotropy in the La$_3$Ni$_2$O$_7$ single crystals with the electric field of the incident light parallel to the crystalline \textit{c}-axis and \textit{ab}-plane, respectively. The evolution of the low-energy part of its \textit{c}-axis optical conductivity spectra ($\mathit{\sigma}_1\mathit{_c}(\mathit{\omega})$) from a Drude component to a finite-energy peak, together with the change in the \textit{c}-axis electron mean-free-path which is distinctly longer than the \textit{c}-axis lattice constant at 10 K but is shorter than the \textit{c}-axis lattice constant at 300 K, demonstrates a crossover from coherent to incoherent interlayer charge dynamics in La$_3$Ni$_2$O$_7$, which is associated with the variation from weak to strong dissipation within its \textit{ab}-plane. In contrast, the Drude component robust in its $\mathit{\sigma}_1\mathit{_{ab}}(\mathit{\omega})$ and the long \textit{ab}-plane electron mean-free-path greater than the \textit{a}-axis and \textit{b}-axis unit-cell lengths manifest the persistence of coherent in-plane charge dynamics from 10 to 300 K. Thus, the charge dynamics in La$_3$Ni$_2$O$_7$ shows a remarkable anisotropy at \textit{high} temperatures. At \textit{low} temperatures, the large values of the ratio between the \textit{ab}-plane and \textit{c}-axis Drude weights and the ratio $\mathit{\sigma}_1\mathit{_{ab}}(\mathit{\omega}\to0)$/$\mathit{\sigma}_1\mathit{_c}(\mathit{\omega}\to0)$ indicate a strong anisotropy of the \textit{low}-temperature charge dynamics in La$_3$Ni$_2$O$_7$.
\end{abstract}

\maketitle
%\tableofcontents
% \section{} 
The discovery of superconductivity in the Ruddlesden-Popper double-layered perovskite nickelate La$_3$Ni$_2$O$_7$ under high pressure, which exhibits the maximal superconducting transition temperature \textit{T$_c$} $\sim$ 80 K \cite{sun_signatures_2023,zhang_high-temperature_2024,hou2023emergence,wang_pressure-induced_2024}, has generated enormous interest in the scientific community \cite{liu_electronic_2024,chen_critical_2023,fan_superconductivity_2024,dong_visualization_2024,wang__normal_2024,qin_high-_2023,yang_possible_2023,huang_impurity_2023,liu_s_2023,jiang__high-temperature_2024,lange_feshbach_2024,qu_bilayer_2024,chen_polymorphism_2024,gao_is_2024,dan2024spin,abadi2024,khasanov2024pressure,sakakibara_possible_2024,takegami_absence_2024,cui_strain-mediated_2024,heier_competing_2024,wu_superexchange_2024,craco_strange_2024,sakakibara_theoretical_2024,huo_modulation_nodate,kakoi_multiband_2024,jiao_enhanced_2024,zhou_evidence_nodate,chen_evidence_2024,lechermann_electronic_2024,geisler_structural_2024}. Considering that (i) the double-layered perovskite nickelate La$_3$Ni$_2$O$_7$, which crystallizes into an orthorhombic phase with a corner-connected NiO$_6$ octahedral layer separated by a La–O layer along the \textit{c}-axis at ambient pressure (see the schematic of the crystal structure in the inset of Fig. 1(a)) and still hosts the layered crystal structure at high pressures although a structure phase transition occurs at pressures above 10 GPa \cite{sun_signatures_2023,geisler_structural_2024,liu2023evidence,puphal_unconventional_2023,wang_long-range_2024,wang_structure_2024,zhang_structural_2024,rhodes_structural_2024}, and (ii) the crystal structures of solids are intimately associated with the electronic properties, studying the discrepancy between the \textit{ab}-plane and \textit{c}-axis electronic properties of La$_3$Ni$_2$O$_7$ is crucial for developing the theory of anisotropic superconductivity in this double-layered perovskite nickelate \cite{fan_superconductivity_2024,gu_effective_2023,yang_orbital-dependent_2024,shilenko_correlated_2023,luo_bilayer_2023,yang_strong_2023,zhang_trends_2023,yang_interlayer_2023,chen_orbital-selective_2024,shen__effective_2023,christiansson_correlated_2023,oh_type-ii_2023,zhang_electronic_2023,liao_electron_2023,lechermann_electronic_2023,zhang_electronic_2024,kaneko_pair_2024,cao_flat_2024,labollita_electronic_2024,chen_electronic_2024,ouyang_hund_2024,jiang_pressure_2024,lu_interlayer-coupling-driven_2024,geisler_optical_2024,tian_correlation_2024,homes_coherence_2005,tajima1997optical,lavrov_normal-state_1998,mandrus1991giant}. However, the anisotropy of the electronic properties in La$_3$Ni$_2$O$_7$ has so far been little investigated by experiments.

Charge dynamics is one of fundamental electronic properties, so studying the anisotropy of the charge dynamics in La$_3$Ni$_2$O$_7$ is an effective approach for gain insight into its electronic anisotropy \cite{homes_coherence_2005,tajima1997optical,dressel2002electrodynamics,basov_electrodynamics_2005,moon_incoherent_2011,uchida_c_1996,uchida_optical_1991,chen_measurement_2010,moon_interlayer_2013,cheng_three-dimensionality_2011,ishikawa_optical_1998,ruzicka_charge_2001,katsufuji_-plane_1996}. For some layered materials, remarkable difference exists between the in-plane and interlayer charge dynamics, e.g., a few underdoped copper oxides and FeTe$_{0.55}$Se$_{0.45}$ show coherent in-plane charge dynamics but incoherent interlayer charge dynamics \cite{,moon_incoherent_2011,uchida_c_1996,uchida_optical_1991}. On the contrary, both in-plane and interlayer charge dynamics are coherent in some other layered materials, such as the undoped iron pnictides BaFe$_2$As$_2$, the optimally doped iron pnictides Ba(Fe$_{0.926}$Co$_{0.074}$)$_2$As$_2$, Ba$_{0.67}$K$_{0.33}$Fe$_2$As$_2$ and the optimally doped or overdoped cuprates \cite{uchida_c_1996,uchida_optical_1991,chen_measurement_2010,moon_interlayer_2013,cheng_three-dimensionality_2011}. Therefore, a natural and important question arises as to what degree of the charge-dynamics anisotropy is in La$_3$Ni$_2$O$_7$, which is expected to provide an essential boundary for understanding the superconducting anisotropy in this double-layered perovskite nickelate \cite{moon_incoherent_2011,uchida_c_1996,uchida_optical_1991,chen_measurement_2010,moon_interlayer_2013,cheng_three-dimensionality_2011,ishikawa_optical_1998,ruzicka_charge_2001,katsufuji_-plane_1996}. However, as one of the prerequisites for obtaining the degree of the charge-dynamics anisotropy, the interlayer charge dynamics in La$_3$Ni$_2$O$_7$ remains elusive.

In underdoped YBa$_2$Cu$_3$O$_y$, YBa$_2$Cu$_4$O$_8$ and overdoped Bi$_2$Sr$_2$CaCu$_2$O$_{8+\delta}$, a crossover from incoherent to coherent interlayer charge dynamics were observed as the temperature decreases \cite{tajima1997optical,lavrov_normal-state_1998,hussey_anisotropic_1997,kaminski_crossover_2003}. The observed incoherent-to-coherent crossover of the interlayer charge dynamics guarantees the three-dimensional metallic states at low temperatures, which was proposed to be significant for the superconducting pair tunnelling along the crystalline \textit{c}-axis and the emergence of the high-\textit{T$_c$} superconductivity in the copper oxides \cite{homes_coherence_2005,tajima1997optical,lavrov_normal-state_1998,mandrus1991giant,kumar_c_1998,chakravarty_interlayer_1993}. Nonetheless, the temperature evolution of the interlayer charge dynamics in La$_3$Ni$_2$O$_7$ which has a similar crystal structure with the copper oxides has so far been little investigated by experiments.

\begin{figure}[b]
	\includegraphics[width=0.98\columnwidth]{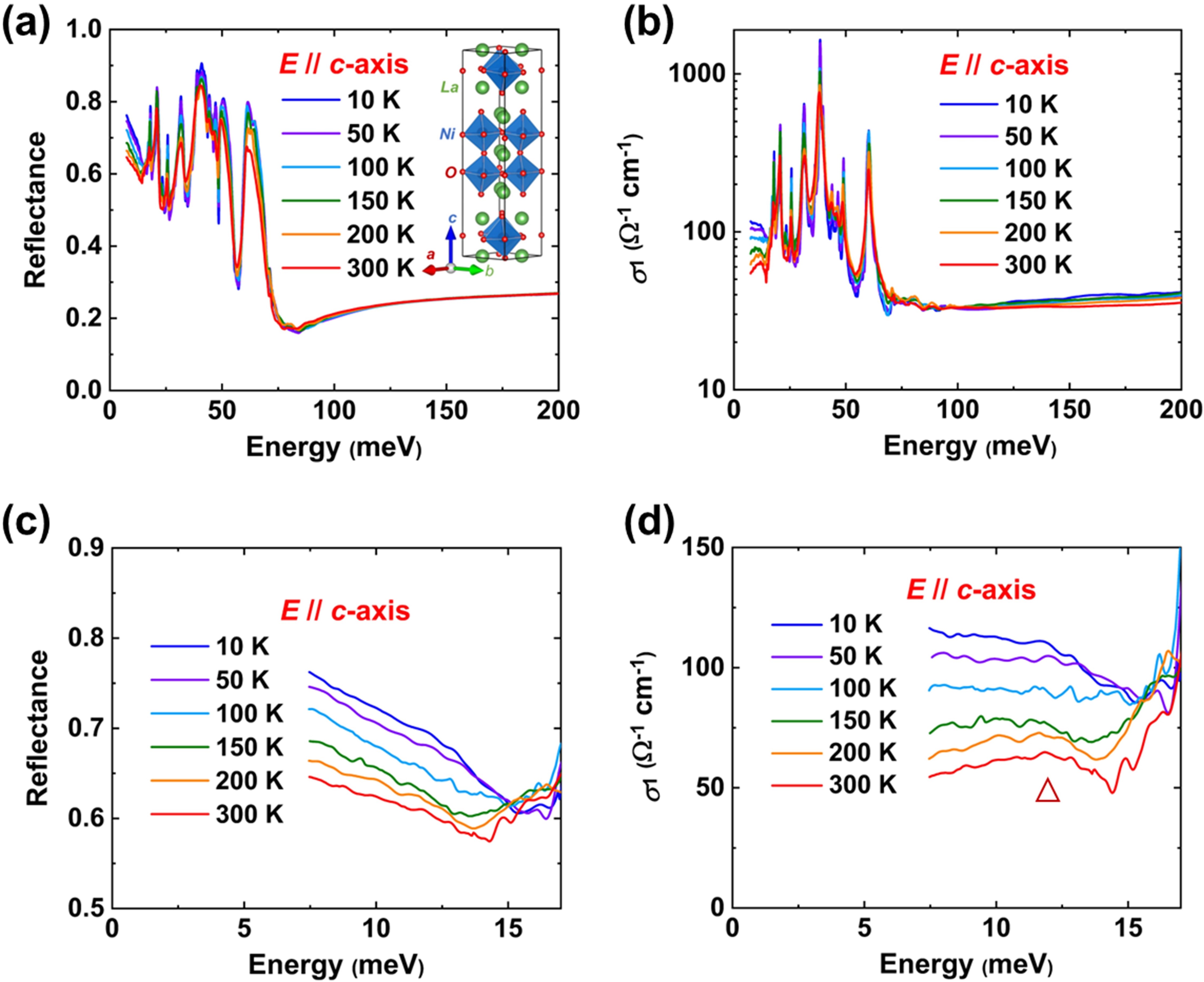}
	\caption{\label{fig1}\textit{C}-axis optical response of La$_3$Ni$_2$O$_7$. (a) \textit{c}-axis reflectance $\mathit{R}_\mathit{c}(\omega)$ below 200 meV. The inset indicates the crystal structure of La$_3$Ni$_2$O$_7$ at ambient pressure. (b) \textit{c}-axis optical conductivity $\mathit{\sigma}_1\mathit{_{c}}(\mathit{\omega})$ below 200 meV. (c) $\mathit{R}_\mathit{c}(\omega)$ up to 17 meV. (d) $\mathit{\sigma}_1\mathit{_{c}}(\mathit{\omega})$ up to 17 meV. The red triangle indicates the finite-energy peak in the low-energy part of the $\mathit{\sigma}_1\mathit{_{c}}(\mathit{\omega})$ at high temperatures.}
\end{figure}

Optical spectroscopy is an efficient experimental technique for studying charge-dynamics anisotropy of a material as it can probe itinerant charge carriers using linearly polarized incident light \cite{dressel2002electrodynamics,basov_electrodynamics_2005,moon_incoherent_2011,uchida_c_1996,uchida_optical_1991,chen_measurement_2010,moon_interlayer_2013,cheng_three-dimensionality_2011,ishikawa_optical_1998,ruzicka_charge_2001,katsufuji_-plane_1996,homes_optical_1993,dordevic_anisotropic_2001,chen_optical_2010,su_strong_2019,takenaka_coherent--incoherent_2002}. Here, to study the charge-dynamics anisotropy in La$_3$Ni$_2$O$_7$, we measured the optical reflectance spectra (i.e.,$\mathit{R}(\mathit{\omega})$) of its single crystals at ambient pressure at low temperatures over a broad photon-energy ($\omega$) range with the light electric field \textit{\textbf{E}} $\parallel$ \textit{ab}-plane and \textit{\textbf{E}} $\parallel$ \textit{c}-axis, respectively (see the details about the reflectance measurements, the sample preparation and the characterizations of the two measured crystal faces which are parallel to the \textit{ab}-plane and \textit{c}-axis, respectively in Supplemental Material). The real part ($\mathit{\sigma}_1\mathit(\mathit{\omega})$) of the optical conductivity is obtained via the Kramers-Kronig transformation of the $\mathit{R}(\mathit{\omega})$. Figure 1(a) depicts the \textit{c}-axis reflectance spectra $\mathit{R}_\mathit{c}(\omega)$ of La$_3$Ni$_2$O$_7$ in the photon-energy range up to 200 meV at different temperatures (see the schematic of the crystal structure in the inset of Fig. 1(a)). Several peak- and dip-like sharp features present in the energy range from 14 to 75 meV in the  $\mathit{R}_\mathit{c}(\omega)$ correspond to the spike-like features in the $\mathit{\sigma}_1\mathit{_{c}}(\mathit{\omega})$ (see Fig. 1(b)), which are assigned as the phonon modes (see the calculated phonon modes in Table S1 and the related calculation details in Supplemental Material). To clearly show the low-energy optical response of electrons, we plotted the  $\mathit{R}_\mathit{c}(\omega)$ and the $\mathit{\sigma}_1\mathit{_{c}}(\mathit{\omega})$ in the energy range up to 17 meV in Fig. 1(c) and 1(d), respectively. Noteworthily, in Fig. 1(d), the $\mathit{\sigma}_1\mathit{_{c}}(\mathit{\omega})$ at 10 K, 50 K and 100 K exhibit the Drude components---upturn-like features arising from the optical response of the itinerant charge carriers at energies lower than 14 meV, which qualitatively indicate that interlayer charge dynamics in La$_3$Ni$_2$O$_7$ is coherent at 10 K, 50 K and 100 K. As the temperature rises, the upturn-like features at 10 K, 50 K and 100 K evolves into the obvious peak-like features centered at finite energies in the low-energy part of the $\mathit{\sigma}_1\mathit{_{c}}(\mathit{\omega})$ at 150 K, 200 K and 300 K (see Fig. 1(d)). The finite-energy peak in the low-energy part of the $\mathit{\sigma}_1\mathit{_{c}}(\mathit{\omega})$ at high temperatures can serve as a spectroscopic signature of incoherent interlayer charge dynamics in La$_3$Ni$_2$O$_7$. Thus, the evolution from the low-temperature Drude feature to the high-temperature finite-energy peak in the low-energy part of the $\mathit{\sigma}_1\mathit{_{c}}(\mathit{\omega})$ here, which was similarly observed in the copper oxides La$_{2-x}$Sr$_x$CuO$_4$ and Bi$_2$Sr$_2$CuO$_6$ \cite{takenaka_coherent--incoherent_2002,tsvetkov_-plane_1997}, suggests that as the temperature increases, a crossover from coherent to incoherent interlayer charge dynamics occurs in La$_3$Ni$_2$O$_7$.

To quantitatively check the temperature-induced crossover from coherent to incoherent interlayer charge dynamics in La$_3$Ni$_2$O$_7$, we need to compare the electron mean free path (\textit{l$_c$}) along the \textit{c}-axis with its \textit{c}-axis lattice constant (\textit{d$_c$}) according to the Mott-Ioffe-Regel criterion \cite{ioffe1960non,mott_conduction_1972,hussey__universality_2004}. In principle, the \textit{c}-axis mean free path \textit{l$_{c}$} can be estimated via the following relationship:
\begin{equation}
l_{c} =v_{F}^{Z}\tau_{c},     
\end{equation} 
where $v_{F}^{Z}$ is the component of the Fermi velocity along \textit{k$_z$} direction, and $\tau_{\text{c}}$ is the relaxation time along the \textit{c}-axis. To obtain the \textit{c}-axis relaxation time $\tau_{\text{c}}$, we fit the $\mathit{\sigma}_1\mathit{_c}(\mathit{\omega})$ at different temperatures using a standard Drude-Lorentz model (see the fitting parameters in Table S2-7, the Drude and Lorentzian components in Fig. S3 of Supplemental Material) \cite{dressel2002electrodynamics,basov_electrodynamics_2005}. Figure 2(a) and 2(b) present the Drude-Lorentz fits to the $\mathit{\sigma}_1\mathit{_c}(\mathit{\omega})$ at two typical temperatures 10 K and 300 K, respectively (see the gray fitting curves and the Drude components shaded in violet color). In Fig. 2(c), the scattering rate (i.e., $\tau_{c}^{-1}$) of the itinerant charge carriers along the \textit{c}-axis, which is equal to the half-height width of the Drude component, decreases from (96.8 $\pm$ 10) to (36.6 $\pm$ 3) meV as the temperature is lowered from 300 to 10 K. Correspondingly, the $\tau_{c}$ increases from (4.3 $\pm$ 0.4) $\times$ 10$^{-14}$ s to (1.1 $\pm$ 0.1) $\times$ 10$^{-13}$ s with decreasing temperature from 300 to 10 K. To evaluate the \textit{c}-axis component of the Fermi velocity, we performed \textit{ab initio} calculations of the Fermi surfaces (FSs) based on the optimized cell geometry at ambient pressure (see the details about \textit{ab initio} calculations in Supplemental Material) \cite{sun_signatures_2023}. In Fig. 2(d), not only the calculated FSs around the $M$ point, the $X$ point and the $Y$ point but also the calculated FSs around the $\Gamma$ point along the $X-\Gamma-X$ direction and the $Y-\Gamma-Y$ direction host near-zero \textit{v}$_{F}^{Z}$, while the calculated FSs around the $\Gamma$ point along the $M-\Gamma-M$ direction has a nonnegligible component of the Fermi velocity along \textit{k$_z$}-direction \textit{v}$_{F(DFT)}^{Z}$ = 6.71 $\times$ 10$^{4}$ m/s. Noteworthily, compared with the measured electronic bands, the electronic bands obtained by \textit{ab initio} calculations host the reduced Fermi velocities due to orbital-dependent electronic correlations \cite{yang_orbital-dependent_2024,shilenko_correlated_2023,liao_electron_2023,lechermann_electronic_2023,cao_flat_2024,ouyang_hund_2024}. Considering the renormalization factor ($\sim$ 1.8) for the electronic band generating the FS around the $\Gamma$ point along the $M-\Gamma-M$ direction \cite{yang_orbital-dependent_2024}, the \textit{k$_z$}-direction component of the renormalized Fermi velocity on the FS around the $\Gamma$ point along the $M-\Gamma-M$ direction \textit{v}$_{F(REN)}^{Z}$ = \textit{v}$_{F(DFT)}^{Z}$/1.8 $\approx$ 3.72 $\times$ 10$^4$ m/s. Thus, according to Eq. (1), the \textit{l$_c$} at 10 K and 300 K can be estimated to be $\sim$ (42.1 $\pm$ 4) \AA\ and $\sim$ (15.9 $\pm$ 2) \AA, respectively. Therein, the \textit{l$_c$} at 10 K are distinctly longer than the \textit{d$_c$} of La$_3$Ni$_2$O$_7$, which quantitatively demonstrates the coherent interlayer charge dynamics in La$_3$Ni$_2$O$_7$ at 10 K. By contrast, the \textit{l$_c$} at 300 K is a little bit shorter than the \textit{d$_c$}, which quantitatively confirms the incoherent interlayer charge dynamics in La$_3$Ni$_2$O$_7$ at 300 K. Figure 2(e) shows that the \textit{l$_c$} at \textit{T} $<$ 150 K is remarkably longer than the \textit{d$_c$}, while the \textit{l$_c$} at \textit{T} $\ge$ 150 K is shorter than the \textit{d$_c$}, which further verifies the temperature-induced crossover from coherent to incoherent interlayer charge dynamics in La$_3$Ni$_2$O$_7$.

\begin{figure}[b]
	\includegraphics[width=0.98\columnwidth]{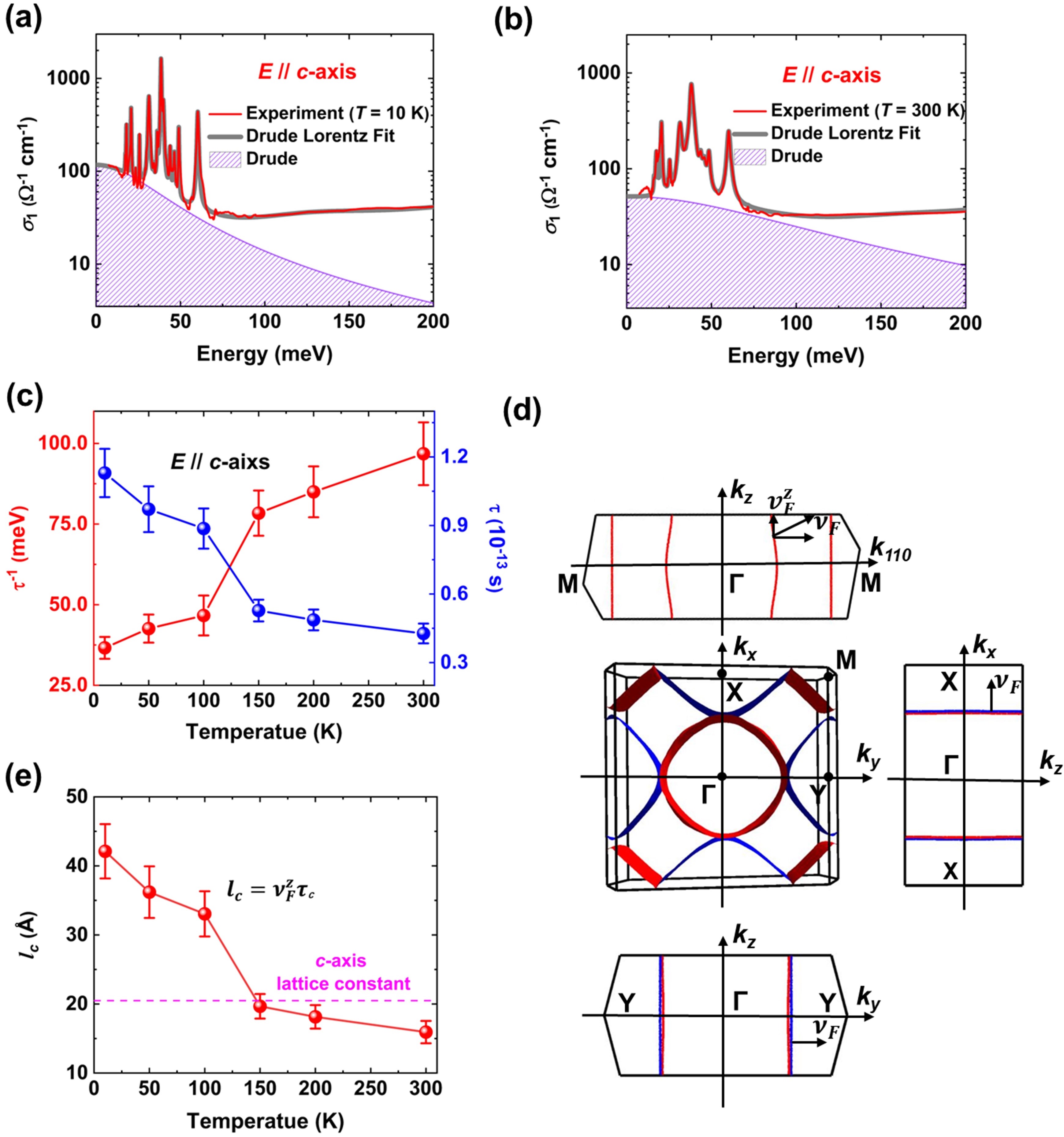}
	\caption{\label{fig2} Estimation of the \textit{c}-axis mean free path \textit{l$_{c}$}. (a) and (b) Drude components in the $\mathit{\sigma}_1\mathit{_{c}}(\mathit{\omega})$ at 10 K and 300 K. (c) Temperature evolutions of the \textit{c}-axis scattering rate $\tau_{c}^{-1}$ and relaxation time $\tau_{c}$. (d) Calculated FSs of La$_3$Ni$_2$O$_7$ at ambient pressure in three-dimensional (3D) Brillouin zone and the side views of the FSs along the $M-\Gamma-M$ direction, the $X-\Gamma-X$ direction and the $Y-\Gamma-Y$ direction. (e) Estimated \textit{l$_{c}$} as a function of temperature. The pink dashed line indicates the \textit{c}-axis lattice constant.}
\end{figure}

Previous investigations suggest that the \textit{c}-axis charge dynamics in a large number of layered materials including the copper oxides, the iron pnictide BaFe$_2$As$_2$ and the iron chalcogenide FeTe$_{0.55}$Se$_{0.45}$ may be associated with the dissipation within the \textit{ab}-plane \cite{moon_incoherent_2011,uchida_optical_1991,dordevic_anisotropic_2001,homes_electronic_2010}. To study whether the temperature-induced crossover from coherent to incoherent interlayer charge dynamics in La$_3$Ni$_2$O$_7$ is related to the change in the dissipation within the \textit{ab}-plane, we measured its \textit{ab}-plane reflectance spectra $R_{ab}(\omega)$. Figure 3(a) depicts the measured $R_{ab}(\omega)$ and $\mathit{\sigma}_1\mathit{_{ab}}(\mathit{\omega})$ of the La$_3$Ni$_2$O$_7$ single crystals at two representative temperatures 10 K and 300 K in the energy range up to 200 meV (see the $R_{ab}(\omega)$ and $\mathit{\sigma}_1\mathit{_{ab}}(\mathit{\omega})$ at the other temperatures in Fig. S4 of Supplemental Material). At energies lower than 50 meV, the $R_{ab}(\omega)$ at 10 K not only is higher than that at 300 K but also approaches to unity, which shows the metal-like optical response. Correspondingly, in Fig. 3(a), the $\mathit{\sigma}_1\mathit{_{ab}}(\mathit{\omega})$ at 10 K and 300 K exhibit the Drude components at low energies. To get access to the \textit{ab}-plane charge dynamics in La$_3$Ni$_2$O$_7$, we fit the $\mathit{\sigma}_1\mathit{_{ab}}(\mathit{\omega})$ at 10 K and 300 K based on the standard Drude-Lorentz model (see the fitting parameters in Table S2-7, the Drude and Lorentzian components in Fig. S5 of Supplementary Material). Figure 3(b) and 3(c) display the Drude-Lorentz fits to the $\mathit{\sigma}_1\mathit{_{ab}}(\mathit{\omega})$ at 10 K and 300 K, respectively (see the gray fitting curves and the Drude components shaded in violet color). The \textit{ab}-plane Drude weight \textit{S$_{ab}$} obtained by the Drude-Lorentz fits were plotted as a function of temperature in Fig. 3(d). Since previous theoretical analysis and \textit{ab} \textit{initio} calculations suggest that at ambient pressure, the electronic bands of La$_3$Ni$_2$O$_7$ crossing the Fermi energy are dominated by the single 3\textit{d}$_{x^2-y^2}$ orbital \cite{sun_signatures_2023,yang_possible_2023,geisler_structural_2024,yang_orbital-dependent_2024,zhang_electronic_2023}, the extended Drude model involving the \textit{S$_{ab}$}, which works under the assumption of a single-band response, was employed to obtain the \textit{ab}-plane scattering rate spectra $\tau_{ab}(\omega)$ of La$_3$Ni$_2$O$_7$. Figure 3(e) shows that in the energy region ($\omega$ $<$ 40 meV) of the Drude response at 10 K, the $\tau_{ab}(\omega)$ at 10 K is in the Landau-Fermi-liquid region, i.e., $\tau_{ab}$($\omega$ $<$ 40 meV, \textit{T} = 10 K) $<$ $\omega$, which suggests weak dissipation within the \textit{ab}-plane at 10 K. Similar to the cases in a large variety of layered materials, the weak dissipation within the \textit{ab}-plane of La$_3$Ni$_2$O$_7$ is likely to correspond to the existence of well-defined quasi-particles and is associated with its coherent \textit{c}-axis charge dynamics. On the contrary, the $\tau_{ab}$($\omega$ $<$ 40 meV) at 300 K is beyond the Landau-Fermi-liquid region, i.e., $\tau_{ab}$($\omega$ $<$ 40 meV, \textit{T} = 300 K) $> \omega$, which suggests strong dissipation within the \textit{ab}-plane at 300 K and is consistent with the incoherent \textit{c}-axis charge dynamics revealed by the finite-energy peak in the low-energy part of the $\mathit{\sigma}_1\mathit{_{c}}(\mathit{\omega})$ and the estimated \textit{l$_c$} shorter than the \textit{d$_c$}. Thus, our results show that the temperature-induced crossover from coherent to incoherent interlayer charge dynamics in La$_3$Ni$_2$O$_7$ correlates to the variation in the dissipation within the \textit{ab}-plane.

\begin{figure}[t]
	\includegraphics[width=0.98\columnwidth]{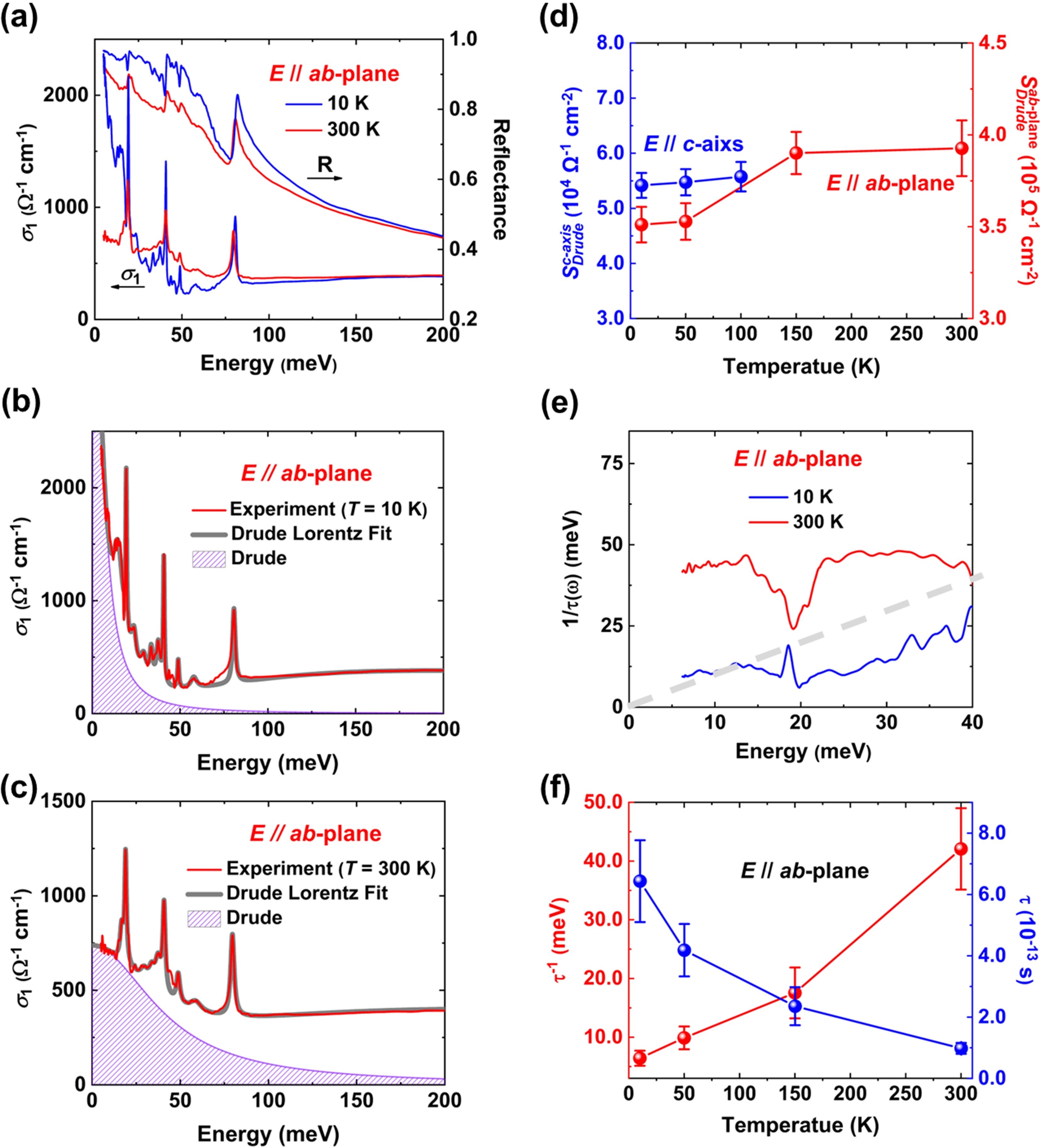}
	\caption{\label{fig3} \textit{Ab}-plane optical response of La$_3$Ni$_2$O$_7$. (a) \textit{ab}-plane reflectance $\mathit{R}_\mathit{ab}(\omega)$ and optical conductivity $\mathit{\sigma}_1\mathit{_{ab}}(\mathit{\omega})$ at 10 K and 300 K in the energy range up to 200 meV. (b) and (c) Drude components in the $\mathit{\sigma}_1\mathit{_{ab}}(\mathit{\omega})$ at 10 K and 300 K. (d) \textit{ab}-plane and \textit{c}-axis Drude weights (\textit{S$_{ab}$} and \textit{S$_{c}$}) at different temperatures. (e) \textit{ab}-plane scattering rate spectra 1/$\tau_{ab}$($\omega$) at 10 K and 300 K. The dashed line represents the relationship 1/$\tau_{ab}$($\omega$) = $\omega$. (f) \textit{ab}-plane scattering rate $\tau_{ab}^{-1}$ and relaxation time $\tau_{ab}$ at different temperatures.}
\end{figure}

Before studying the degree of the charge-dynamic anisotropy in La$_3$Ni$_2$O$_7$, it is essential to gain insight into the nature of its \textit{ab}-plane charge dynamics in the temperature range from 300 to 10 K. In Fig. 3(f), the \textit{ab}-plane scattering rate at 10 K and 300 K, which were obtained by the Drude-Lorentz fits to the $\mathit{\sigma}_1\mathit{_{ab}}(\mathit{\omega})$, correspond to the \textit{ab}-plane-relaxation times $\tau_{ab}$(10 K) = (6.4 $\pm$ 1.3) $\times$ 10$^{-13}$ s and $\tau_{ab}$(300 K) = (9.8 $\pm$ 1.6) $\times$ 10$^{-14}$ s. Moreover, \textit{ab} \textit{initio} calculations reveal that the FSs around the $\Gamma$ point along the $M-\Gamma-M$ direction, the FSs around the $M$ point along the $M-\Gamma-M$ direction, and the FSs around the $X$ (or $Y$) point along the $X-\Gamma-X$ (or $Y-\Gamma-$Y) direction host the components of the Fermi velocities parallel to the  $k_x$-$k_y$ plane \textit{v}$_{F(DFT)}^{xy}$ $\approx$ 4.70 $\times$ 10$^5$ m/s, 5.00 $\times$ 10$^5$ m/s and 6.06 $\times$ 10$^5$ m/s, respectively (see Fig. 2(d)). Considering that the renormalization factors (i.e., 1.8, 2.6 and 2.3) for the corresponding electronic bands \cite{yang_orbital-dependent_2024}, the experimental $k_x$-$k_y$-plane components of the renormalized Fermi velocities on the FSs around the $\Gamma$ point along the $M-\Gamma-M$ direction, on the FSs around the M point along the $M-\Gamma-M$ direction, and on the FSs around the $X$ (or $Y$) point along the $X-\Gamma-X$ (or $Y-\Gamma-Y$) direction \textit{v}$_{F(REN)}^{xy}$ $\approx$ 2.61 $\times$ 10$^5$ m/s, 1.92 $\times$ 10$^5$ m/s and 2.63 $\times$ 10$^5$ m/s, respectively. Given the relationship among the \textit{ab}-plane mean free paths \textit{l$_{ab}$}, the component of the Fermi velocities parallel to the $k_x$-$k_y$ plane \textit{v}$_{F}^{xy}$ and $\tau_{ab}$:
\begin{equation}
	l_{ab} =v_{F}^{xy}\tau_{ab},     
\end{equation}
for the charge carriers on the FSs around the $\Gamma$ point along the $M-\Gamma-M$ direction, on the FSs around the $M$ point along the $M-\Gamma-M$ direction, and on the FSs around the $X$ (or $Y$) point along the $X-\Gamma-X$ (or $Y-\Gamma-Y$) direction, the \textit{l$_{ab}$}(300 K) $\approx$ (256.6 $\pm$ 40) \AA, (189.1 $\pm$ 30) \AA \ and (258.9 $\pm$ 40) \AA, respectively (see Fig. 4(a)). As the temperature is lowered down to 10 K, the \textit{l$_{ab}$} increase to $\sim$ (1679.4 $\pm$ 350) \AA, $\sim$ (1237.6 $\pm$ 260) \AA\ and $\sim$ (1694.6 $\pm$ 350) \AA, respectively (see Fig. 4(a)). Thus, in the temperature range 300---10 K, all the estimated \textit{l$_{ab}$} are much longer than the unit cell lengths of La$_3$Ni$_2$O$_7$ along the \textit{a}-axis and \textit{b}-axis, i.e., \textit{d$_a$} $\sim$ 5.40 \AA \ and \textit{d$_b$} $\sim$ 5.43 \AA. According to the Mott-Ioffe-Regel criterion, the \textit{ab}-plane charge dynamics in La$_3$Ni$_2$O$_7$ should be coherent from 300 to 10 K, which is quite different from its incoherence interlayer charge dynamics at high temperatures. Thus, La$_3$Ni$_2$O$_7$ exhibits strongly anisotropic charge dynamics at \textit{high} temperatures.
\begin{figure}[t]
	\includegraphics[width=0.98\columnwidth]{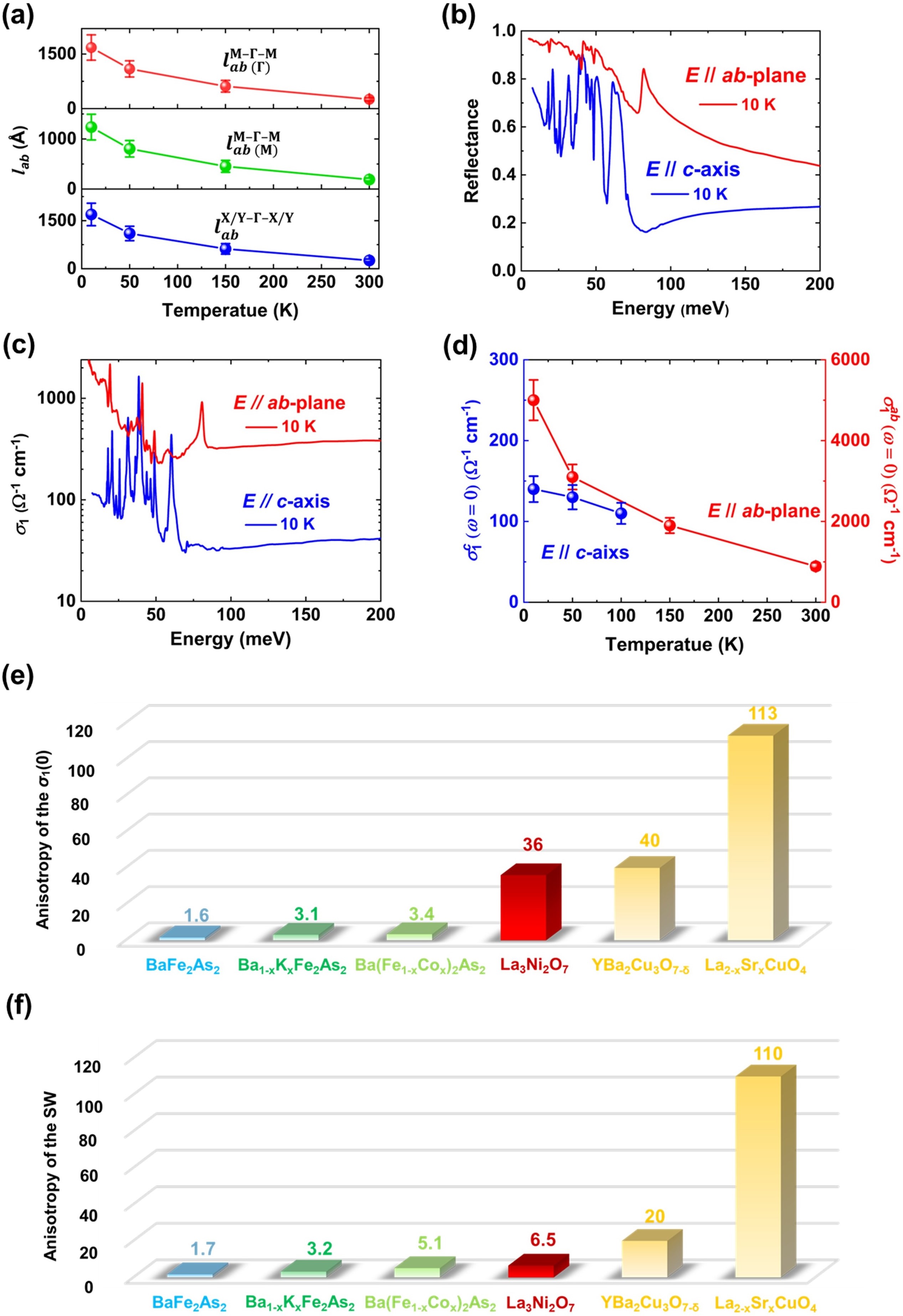}
	\caption{\label{fig4} Strongly anisotropic charge dynamics. (a) \textit{ab}-plane mean free paths \textit{l$_{ab}$} for the charge carriers on the FSs around the $\Gamma$ point along the $M-\Gamma-M$ direction (see the red dots), on the FSs around the $M$ point along the $M-\Gamma-M$ direction (see the green dots) and on the FSs around the $X$ (or $Y$) point along the $X-\Gamma-X$ (or $Y-\Gamma-Y$) direction (see the blue dots). (b) $\mathit{R}_\mathit{ab}(\omega)$ and $\mathit{R}_\mathit{c}(\omega)$ at 10 K. (c) $\mathit{\sigma}_1\mathit{_{ab}}(\mathit{\omega})$ and $\mathit{\sigma}_1\mathit{_{c}}(\mathit{\omega})$ at 10 K. (d) $\mathit{\sigma}_1\mathit{_{ab}}(\mathit{\omega}\to0)$ and $\mathit{\sigma}_1\mathit{_c}(\mathit{\omega}\to0)$ obtained by the Hagen-Rubens low-energy extrapolations. (e) Ratios  $\mathit{\sigma}_1\mathit{_{ab}}(\mathit{\omega}\to0)$/$\mathit{\sigma}_1\mathit{_c}(\mathit{\omega}\to0)$ of La$_3$Ni$_2$O$_7$, the iron-pnictide superconductors and the copper-oxide superconductors at low temperatures \cite{uchida_c_1996,uchida_optical_1991,chen_measurement_2010,moon_interlayer_2013,cheng_three-dimensionality_2011,hu_origin_2008,schutzmann_c_1994,schutzmann_far-infrared_1992}. (f) Ratios \textit{S$_{ab}$}/\textit{S$_{c}$} of La$_3$Ni$_2$O$_7$, the iron-pnictide superconductors and the copper-oxide superconductors at low temperatures \cite{uchida_c_1996,chen_measurement_2010,moon_interlayer_2013,cheng_three-dimensionality_2011,schutzmann_c_1994}. }
\end{figure}

To qualitatively study the charge-dynamic anisotropy in La$_3$Ni$_2$O$_7$ at \textit{low} temperature, we plotted the $\mathit{R}_\mathit{ab}(\omega)$ and  $\mathit{R}_\mathit{c}(\omega)$ at 10 K over a broad energy range up to 200 meV in Fig. 4(b). Compared with the $\mathit{R}_\mathit{c}(\omega)$, the $\mathit{R}_\mathit{ab}(\omega)$ exhibits a very different photon-energy dependence. Correspondingly, Fig. 4(c) shows that the low-energy part ($\omega$ $<$ 50 meV) of the $\mathit{\sigma}_1\mathit{_{ab}}(\mathit{\omega})$ at 10 K, which includes the Drude component, is much higher than that of the $\mathit{\sigma}_1\mathit{_{c}}(\mathit{\omega})$ at 10 K. The great discrepancy between the low-energy part of the $\mathit{\sigma}_1\mathit{_{ab}}(\mathit{\omega})$ and the $\mathit{\sigma}_1\mathit{_{c}}(\mathit{\omega})$ of La$_3$Ni$_2$O$_7$ at 10 K, which reflects the 2D-like FSs, suggests that the anisotropy of its charge dynamics should be also strong at \textit{low} temperatures. Moreover, in Fig. 2(d), our \textit{ab} \textit{initio} calculations show that the \textit{k$_z$} components of all the Fermi velocities normal to the FSs are much smaller than the components of the Fermi velocities parallel to the $k_x$-$k_y$ plane, which indicates two-dimensional (2D) like FSs and further supports the strongly anisotropic charge dynamics in La$_3$Ni$_2$O$_7$ at \textit{low} temperatures. 

To quantitatively study the degree of the charge-dynamics anisotropy in La$_3$Ni$_2$O$_7$ at \textit{low} temperatures, we plotted the  $\mathit{\sigma}_1\mathit{_{ab}}(\mathit{\omega}\to0)$ and the $\mathit{\sigma}_1\mathit{_c}(\mathit{\omega}\to0)$ at different temperatures in Fig. 4(d), which were obtained by the Hagen-Rubens low-energy extrapolation of the $\mathit{R}_\mathit{ab}(\omega)$ and $\mathit{R}_\mathit{c}(\omega)$, respectively (see the Method, Fig. S2 and Fig. S4 in Supplemental Material). At 10 K, the ratio $\mathit{\sigma}_1\mathit{_{ab}}(\mathit{\omega}\to0)$/$\mathit{\sigma}_1\mathit{_c}(\mathit{\omega}\to0)$ of La$_3$Ni$_2$O$_7$ shows the value of $\sim$ 36, which is higher than $\mathit{\sigma}_1\mathit{_{ab}}(\mathit{\omega}\to0)$/$\mathit{\sigma}_1\mathit{_c}(\mathit{\omega}\to0)$ of the ``122'' type iron-pnictide superconductors BaFe$_{2}$As$_{2}$ ($\sim$ 1.6) \cite{chen_measurement_2010,hu_origin_2008}, optimally doped Ba$_{1-x}$K$_{x}$Fe$_{2}$As$_{2}$ ($\sim$ 3.1) and Ba(Fe$_{1-x}$Co$_{x}$)$_{2}$As$_{2}$ ($\sim$ 3.4) \cite{moon_interlayer_2013,cheng_three-dimensionality_2011}, but is significantly less than the $\mathit{\sigma}_1\mathit{_{ab}}(\mathit{\omega}\to0)$/$\mathit{\sigma}_1\mathit{_c}(\mathit{\omega}\to0)$ of the fully oxygenated copper-oxide superconductor YBa$_{2}$Cu$_{3}$O$_{7-\delta}$ ($\sim$ 40) and 
overdoped copper-oxide superconductor La$_{2-x}$Sr$_{x}$CuO$_{4}$ ($\sim$ 113) (see Fig. 4(e)) \cite{uchida_c_1996,uchida_optical_1991,schutzmann_c_1994,schutzmann_far-infrared_1992}. Moreover, according to the \textit{S$_{ab}$} and \textit{S$_{c}$} of La$_3$Ni$_2$O$_7$ in Fig. 3(d), we can get the ratio \textit{S$_{ab}$}/\textit{S$_{c}$} $\sim$ 6.5 at 10 K, which is larger than the \textit{S$_{ab}$}/\textit{S$_{c}$} of the ``122'' type iron-pnictide superconductors BaFe$_{2}$As$_{2}$ ($\sim$ 1.7) \cite{chen_measurement_2010}, optimally doped Ba$_{1-x}$K$_{x}$F$e_{2}$As$_{2}$ ($\sim$ 3.2) and Ba(Fe$_{1-x}$Co$_{x}$)$_{2}$As$_{2}$ ($\sim$ 5.1) \cite{moon_interlayer_2013,cheng_three-dimensionality_2011}, but is much smaller than the \textit{S$_{ab}$}/\textit{S$_{c}$} of the fully oxygenated copper-oxide superconductor YBa$_{2}$Cu$_{3}$O$_{7-\delta}$ ($\sim$ 20) and overdoped copper-oxide superconductor La$_{2-x}$Sr$_{x}$CuO$_{4}$ ($\sim$ 110) (see Fig. 4(f)) \cite{uchida_c_1996,schutzmann_c_1994}. Thus, the large values of the ratios \textit{S$_{ab}$}/\textit{S$_{c}$} and $\mathit{\sigma}_1\mathit{_{ab}}(\mathit{\omega}\to0)$/$\mathit{\sigma}_1\mathit{_c}(\mathit{\omega}\to0)$ in La$_3$Ni$_2$O$_7$ at 10 K indicate that its low-temperature charge dynamics exhibits a strong anisotropy with its degree located between the ``122'' type iron-pnictide superconductors and the fully oxygenated/overdoped copper-oxide superconductors. Noteworthily, the ``122'' type iron-pnictide superconductors host the large-size 3D ellipsoid FS with the dominant Fe-3\textit{d}$_{3z^{2}-r^{2}}$ orbital enclosing the \textit{Z} point of the Brillouin zone and the quasi-2D cylinder-like FSs with small warping along \textit{k$_{z}$} around the $\Gamma$ point and the corners of the Brillouin zone \cite{chen_measurement_2010,liu_k-doping_2008,utfeld_bulk_2010,wang_gutzwiller_2010}, while in La$_3$Ni$_2$O$_7$, except that the FS around the $\Gamma$ point along the $M-\Gamma-M$ direction has a small Fermi-velocity component along\textit{ k$_{z}$} direction, the majority of the FSs are quasi-2D and have the Fermi velocities substantially parallel to the $k_x$-$k_y$ plane (see Fig. 2(d)), which supports that La$_3$Ni$_2$O$_7$ exhibits a stronger anisotropy of charge dynamics than the ``122'' type iron-pnictide superconductors. Moreover, although both the copper oxides and La$_3$Ni$_2$O$_7$ have quasi-2D FSs dominated by 3\textit{d}$_{x^{2}-y^{2}}$ orbitals \cite{sun_signatures_2023,fan_superconductivity_2024,heier_competing_2024,geisler_structural_2024,damascelli_angle-resolved_2003}, a theoretical study indicates that the interlayer hopping term related to the Ni 3\textit{d}$_{x^{2}-y^{2}}$ orbital in La$_3$Ni$_2$O$_7$ is remarkably larger than that associated with the Cu 3\textit{d}$_{x^{2}-y^{2}}$ orbital in the copper oxides \cite{fan_superconductivity_2024,zhang_electronic_2023}, which also suggests that the charge-dynamics anisotropy in La$_3$Ni$_2$O$_7$ is weaker than those in the copper oxides.

In summary, we have investigated the interlayer charge dynamics and the charge-dynamics anisotropy in La$_3$Ni$_2$O$_7$. Considering that (i) as the temperature increases, the Drude component in the low-energy region of the $\mathit{\sigma}_1\mathit{_{c}}(\mathit{\omega})$ evolves into a finite-energy peak, and (ii) at 10 K, \textit{l$_c$} $>$ \textit{d$_c$} but at 300 K, \textit{l$_c$} $<$ \textit{d$_c$}, a crossover from coherent to incoherent interlayer charge dynamics occurs in La$_3$Ni$_2$O$_7$, which is related to the variation from weak to strong dissipation within its \textit{ab}-plane. By contrast, in the temperature range 10---300 K, the robustness of the Drude component in its $\mathit{\sigma}_1\mathit{_{ab}}(\mathit{\omega})$, together with \textit{l$_{ab}$} $\gg$ \textit{d$_a$} and \textit{l$_{ab}$} $\gg$ \textit{d$_b$}, demonstrates that the in-plane charge dynamics is coherent. Thus, the anisotropy of the charge dynamics in La$_3$Ni$_2$O$_7$ is remarkable  at \textit{high} temperatures. At \textit{low} temperatures, the large values of the ratios \textit{S$_{ab}$}/\textit{S$_{c}$} and $\mathit{\sigma}_1\mathit{_{ab}}(\mathit{\omega}\to0)$/$\mathit{\sigma}_1\mathit{_c}(\mathit{\omega}\to0)$ indicate that the anisotropy of the \textit{low}-temperature charge dynamics is strong in La$_3$Ni$_2$O$_7$.

\begin{acknowledgments}
We thanks Prof. Tao Xiang for very helpful and fruitful discussions. The authors acknowledge support from the strategic Priority Research Program of Chinese Academy of Sciences (Project No. XDB33000000), the National Natural Science Foundation of China (Grant No. U21A20432 and 12174454), the National Key Research and Development Program of China (Grant No. 2022YFA1403800 and 2023YFA1406500), the Guangdong Basic and Applied Basic Research Foundation (Projects No. 2021B1515130007 and 2024B1515020040), Guangzhou Basic and Applied Basic Research Funds (Grant No. 2024A04J6417), and Guangdong Provincial Key Laboratory of Magnetoelectric Physics and Devices (Grant No. 2022B1212010008) and the Synergetic Extreme Condition User Facility (SECUF) Infrared Unit in THz and Infrared Experimental Station. Z.-G.C. conceived and supervised this project. C.H., M.H., and M.W. grew the single crystals. C.H., M.H. and B.S. performed the characterization of the single crystals. B.S. carried out the optical experiments. J.Z. did first-principle calculations. Z-G.C., B.S., J.Z. and J.L. analyzed the data. Z-G.C. and B.S. wrote the paper.

$^{\dagger}$B. S., J.Z. and C. H. contributed equally to this work.

\end{acknowledgments}

\appendix


\end{document}